\documentclass{aa}

\usepackage[dvips]{graphicx}
\graphicspath{{./}}

\begin{document}

\thesaurus{02.05.2 ; 08.14.1}

%%%%%%%%%%%%%%%%%%%%%%%%%%%%%%%%%%%%%%%%%%%%%%%%%%%%%%%%%%%%%%%%%%%%%%%%%
\title{Microscopic nuclear equation of state with three-body forces 
      and neutron star structure}
%%%%%%%%%%%%%%%%%%%%%%%%%%%%%%%%%%%%%%%%%%%%%%%%%%%%%%%%%%%%%%%%%%%%%%%%%
\author{M.~Baldo\inst{1}
\and I.~Bombaci\inst{2}
\and G.~F.~Burgio\inst{1}}
%%%%%%%%%%%%%%%%%%%%%%%%%%%%%%%%%%%%%%%%%%%%%%%%%%%%%%%%%%%%%%%%%%%%%%%%%
\institute{Dipartimento di Fisica, Universit\'a
 di Catania and I.N.F.N. Sezione di Catania, c.so Italia 57, 
I-95129 Catania, Italy
\and Dipartimento di Fisica, Universit\'a di Pisa and 
     I.N.F.N. Sezione di Pisa, Piazza Torricelli 2, I-56100 Pisa, Italy}
%%%%%%%%%%%%%%%%%%%%%%%%%%%%%%%%%%%%%%%%%%%%%%%%%%%%%%%%%%%%%%%%%%%%%%%%%

%\offprints{G.~F.~Burgio}

\maketitle

%%%%%%%%%%%%%%%%%%%%%%%%%%%%%%%%%%%%%%%%%%%%%%%%%%%%%%%%%%%%%%%%%%%%%%%%%
\begin{abstract}
We calculate static properties of non-rotating neutron stars (NS's) using  
a microscopic equation of state (EOS) for asymmetric nuclear matter,
derived from the Brueckner--Bethe--Goldstone many--body theory with  
explicit three-body forces.
We  use the Argonne AV14 and the Paris two--body nuclear force, 
implemented by the Urbana model for the three-body force.  
We obtain a maximum mass configuration with $ M_{max} = 1.8 M_{\sun}$ 
($M_{max} = 1.94 M_{\sun}$) when the AV14 (Paris) interaction is used.  
They are both consistent with the observed range of NS masses.
The onset of direct Urca processes occurs at densities  
$n \geq 0.65~fm^{-3}$ for the AV14 potential and $n \geq 0.54~fm^{-3}$ 
for the Paris potential. 
Therefore, NS's with masses above $M^{Urca} = 1.4 M_{\sun}$
for the AV14 and $M^{Urca} = 1.24 M_{\sun}$ for the Paris potential
can undergo very rapid cooling, depending on the strength of 
superfluidity in the interior of the NS.  
The comparison with other microscopic models for the EOS shows noticeable
differences.


\end{abstract}
%%%%%%%%%%%%%%%%%%%%%%%%%%%%%%%%%%%%%%%%%%%%%%%%%%%%%%%%%%%%%%%%%%%%%%%%%
\section{Introduction }
%%%%%%%%%%%%%%%%%%%%%%%%%%%%%%%%%%%%%%%%%%%%%%%%%%%%%%%%%%%%%%%%%%%%%%%%%
In the next few years it is expected that a large amount of novel 
informations on neutron stars (NS's) will be available from the 
new generation of X--ray and $\gamma$--ray satellites. 
Therefore, a great interest is devoted presently to the study of 
NS's and to the prediction of their structure on the basis of the 
properties of dense matter. 
The equation of state (EOS) of NS matter covers a wide density range, 
from $\sim 10$ g/cm$^3$ in the surface to several times nuclear matter 
saturation density ($\rho_0 \sim 2.8~10^{14}$ g/cm$^3$) in the center 
of the star (Shapiro et al. 1983). 
The interior part (core) of a NS is made by asymmetric nuclear matter 
with a certain lepton fraction.  
At ultra--high density,  matter might suffer a transition to other
exotic hadronic components (like hyperons, a $K^-$ condensate or a 
deconfined phase of quark matter). The possible appearance of such an 
exotic core has enormous  consequences for the neutron star and black hole 
formation mechanism (Bombaci 1996).    
Unfortunately large uncertainities are still present in the theoretical 
treatment of this ultra--dense regime (Glendenning 1985, Prakash et al.
1997). 
Therefore, in the present work, we consider a more conventional picture 
assuming the NS core is composed only by an uncharged mixture of neutrons, 
protons, electrons and muons in equilibrium with respect to the 
weak interaction  ($\beta$--stable matter).  
Even in this picture, the determination of the EOS of asymmetric nuclear 
matter to describe the core of the NS, remains a  formidable theoretical 
problem (Hjorth-Jensen et al. 1995).  

Any ``realistic'' EOS must satisfy several requirements :
i) It must display the correct saturation point for symmetric
nuclear matter (SNM); ii) it must give a symmetry energy 
compatible with nuclear phenomenology and well behaved at high
densities; iii) for SNM the incompressibility at saturation must
be compatible with the values extracted from phenomenology
(Myers et al. 1996); 
iv) both for neutron matter (NEM) and SNM the speed of sound must
not exceed the speed of light (causality condition), at least up to the
relevant densities; the latter condition is automatically satisfied only
in fully relativistic theory.

In this work we present results for some NS properties obtained on the 
basis of a microscopic EOS which satisfies 
requirements i-iv, and compare them with the predictions of other 
microscopic EOS's. 

%%%%%%%%%%%%%%%%%%%%%%%%%%%%%%%%%%%%%%%%%%%%%%%%%%%%%%%%%%%%%%%%%%%%%%%%%
           \section{Equation of state }
%%%%%%%%%%%%%%%%%%%%%%%%%%%%%%%%%%%%%%%%%%%%%%%%%%%%%%%%%%%%%%%%%%%%%%%%%
\subsection{ Brueckner--Bethe--Goldstone theory  }
%%%%%%%%%%%%%%%%%%%%%%%%%%%%%%%%%%%%%%%%%%%%%%%%%%%
 The Brueckner--Bethe--Goldstone (BBG) theory is based on a linked cluster 
expansion of the energy per nucleon of nuclear matter (see {\it e.g.} 
Bethe 1971). 
The basic ingredient in this many--body approach is the Brueckner reaction 
matrix $G$, which is the solution of the  Bethe--Goldstone equation 

\begin{equation}
G(n;\omega) = v  + v \sum_{k_a k_b} {{|k_a k_b\rangle  Q  \langle k_a k_b|}
  \over {\omega - e(k_a) - e(k_b) }} G(n;\omega), 
\end{equation}                                                           
\noindent
where $v$ is the bare nucleon-nucleon (NN) interaction, $n$ is the nucleon 
number density, and $\omega$ the  starting energy.  
The operator  $|k_a k_b\rangle Q \langle k_a k_b|$ projects on intermediate 
scattering states in which both nucleons are above the Fermi sea 
(Pauli operator).  

\begin{equation}
e(k) = e(k;n) = {{\hbar^2}\over {2m}}k^2 + U(k;n)
\end{equation}
\noindent
is the single particle energy.  
The Brueckner--Hartree--Fock (BHF) approximation for the s.p. potential
$U(k;n)$  using the  {\it continuous choice} is

\begin{equation}
U(k;n) = \sum _{k'\leq k_F} \langle k k'|G(n; e(k)+e(k'))|k k'\rangle_a 
\end{equation}
\noindent
where the subscript ``{\it a}'' indicates antisymmetrization of the 
matrix element.  In this approach equations (1)--(3) have to be solved 
selfconsistently. In the BHF approximation the energy per nucleon is:
 
\begin{equation}
{E \over{A}}  =  
          {{3}\over{5}}{{\hbar^2}\over {2m}}k_F^2  + D_{BHF} ~,
\end{equation}

\begin{equation}
D_{BHF}(n) = {{1}\over{2}}  
{{1}\over{A}}~ \sum_{k,k'\leq k_F} \langle k k'|G(n; e(k)+e(k'))|k k'\rangle_a 
\end{equation}
\noindent
In this scheme, the only input quantity we need is the bare NN interaction
$v$ in the Bethe-Goldstone equation (1). In this sense the BBG 
approach can be considered as a microscopic one.   

%%%%%%%%%%%%%%%%%%%%%%%%%%%%%%%%%%%%%%%%%%%%%%%%%%%%%%%%%
% Figure 1
\begin{figure} [h]
 \begin{center}
\includegraphics[bb= 60 0 515 669,angle=90,scale=0.6]{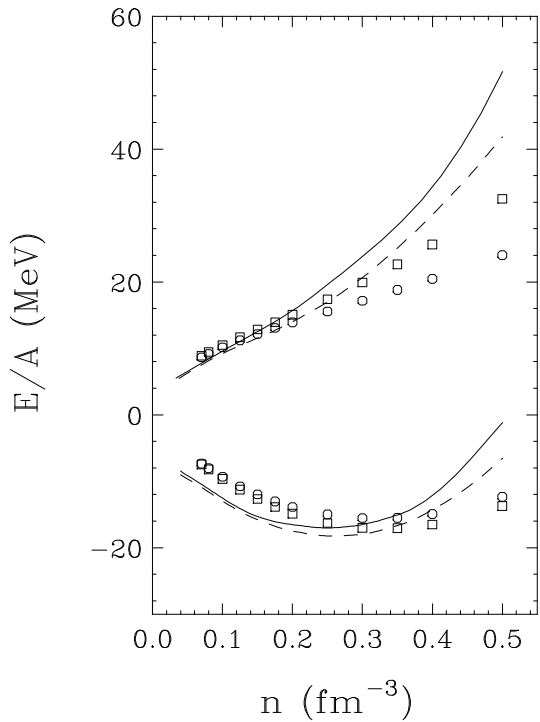}
\end{center}
   \caption{The energy per baryon E/A is plotted vs. the 
number density n  for symmetric matter (lower curves) and 
for neutron matter (upper curves). The solid (dashed) line represents
a non-relativistic Brueckner calculation with Paris (AV14) potential, 
whereas the open circles (squares) are the results of a variational 
calculation performed with AV14 (UV14) interaction.}
    \label{Fig1}
\end{figure}
%%%%%%%%%%%%%%%%%%%%%%%%%%%%%%%%%%%%%%%%%%%%%%%%%%%%%%%%%

The Brueckner-Hartree-Fock (BHF) approximation for the EOS in SNM, within the 
continuous choice (Baldo et al. 1991), reproduces closely results which 
include up to four hole line diagram contributions to the BBG expansion of 
the energy, calculated within the so called {\it gap choice} for the single 
particle potential $U(k)$ (Day et al. 1985). 
In the following we refer to the latter calculations as  $BBG_4^{gap}$. 
The numerical accuracy of the solution of the BHF equation in the 
continuous choice has been discussed recently in ref. (Schulze et al. 1995).

In Fig. 1 we show the energy per nucleon calculated within this scheme 
in the case of SNM (lower curves) and NEM (upper curves);  
the solid and dashed curves have been obtained using respectively the 
Paris potential (Lacombe et al. 1980) and the Argonne $v_{14}$ (AV14) model 
(Wiringa et al. 1984) for the two-body nuclear force.   
For comparison, we report in the same figure the energy per nucleon 
calculated within the variational many--body approach 
(Wiringa et al. 1988, hereafter WFF) using respectively the 
AV14 (open circles) and the Urbana UV14 (squares)   
two-nucleon potentials.

We notice that both the BHF and the variational approach fail
to reproduce the empirical saturation point of nuclear matter
($n_o = 0.17 \pm 0.01~fm^{-3}$, $E_o/A = -16 \pm 1~MeV$).
Analogous non-relativistic calculations have been performed with the 
Bonn potential (Machleidt 1989, Engvik et al. 1996). Even in that case 
symmetric nuclear matter does not saturate at the empirical point.

The comparison of the saturation curves obtained in the BHF and
the variational approach with the same two-nucleon
AV14 potential shows that our BHF saturation point
($n_o = 0.256~fm^{-3}$, $E_o/A = -18.26~MeV$) 
is closer to the empirical saturation point than  
the one reported in WFF paper 
($n_o = 0.319~fm^{-3}$, $E_o/A = -15.6~MeV$). 
The two methods give results in reasonable agreement 
in the case of SNM up to a density $n \simeq 0.4~fm^{-3}$, whereas  
sizeable differences are evident for NEM already at densities 
$n \simeq 0.3~fm^{-3}$. A similar outcome has been found by comparing 
variational and $BBG_4^{gap}$ saturation curves for the same NN potential 
(Day, et al 1985). The discrepancy clearly arises from the different 
many--body technique.

In Fig.2 we report the energy per nucleon as a function of the number density
$n$ calculated in the BHF approximation using the Paris (solid curves)
and the AV14 potentials (dashed curves) up to densities typically
encountered in the core of a neutron star. The upper curves refer to 
neutron matter and the lower ones to symmetric nuclear matter.
We notice that the EOS derived with the Paris potential is stiffer than the one
obtained with the AV14 interaction. This is possibly due to the 
momentum dependence of the Paris potential, as discussed recently in 
ref.(Engvik et al. 1997).

%%%%%%%%%%%%%%%%%%%%%%%%%%%%%%%%%%%%%%%%%%%%%%%%%%%%%%%%%
% Figure 2
\begin{figure} [h]
 \begin{center}
\includegraphics[bb= 60 0 515 669,angle=90,scale=0.6]{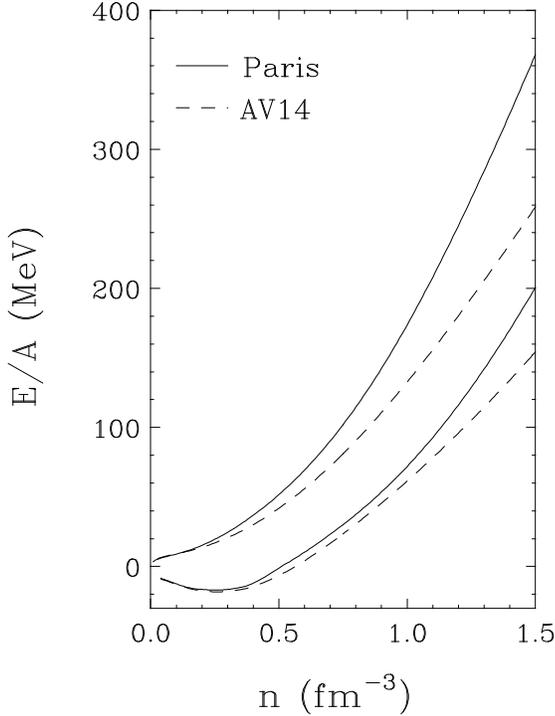}
\end{center}
\caption
{The energy per baryon E/A is plotted vs. the 
number density n  for symmetric matter (lower curves) and 
for neutron matter (upper curves). 
Non-relativistic Brueckner calculations are drawn for the AV14 
(dashed line) and the Paris (solid line) potentials.}
    \label{Fig2}
\end{figure}
%%%%%%%%%%%%%%%%%%%%%%%%%%%%%%%%%%%%%%%%%%%%%%%%%%%%%%%%%

%%%%%%%%%%%%%%%%%%%%%%%%%%%%%%%%%%%%%%%%%%%%%%%%%%%
\subsection{Three-body forces}
%%%%%%%%%%%%%%%%%%%%%%%%%%%%%%%%%%%%%%%%%%%%%%%%%%%
Non--relativistic calculations, based on purely two--body interactions, fail 
to reproduce the correct saturation point for symmetric nuclear matter 
(Coester et al. 1970). 
This well known deficiency is commonly corrected introducing 
three-body forces (TBF). 
Unfortunately, it seems not possible to reproduce the experimental binding 
energies of light nuclei and the correct saturation point accurately with one  
simple set of  TBF (Wiringa et al. 1988). 
Relevant progress has been made in the
theory of nucleon TBF, but a complete theory is not yet available. 
A realistic model for nuclear TBF has been introduced by the Urbana group 
(Carlson et al. 1983; Schiavilla et al. 1986).  
The Urbana model consists of an attractive term  $V^{2\pi}_{ijk}$  due to 
two--pion exchange with excitation of an intermediate $\Delta$-resonance, 
and a repulsive phenomenological central term $V^{R}_{ijk}$

\begin{equation} 
V_{ijk} =  V^{2\pi}_{ijk} +  V^{R}_{ijk} 
\end{equation}
\noindent
The two--pion exchange contribution is a cyclic sum over the nucleon indices 
{\it i, j, k} of products of anticommutator \{,\} and commutator [,] terms

\begin{eqnarray} 
V^{2\pi}_{ijk} & = &  A \sum_{cyc} \Big( \{X_{ij},X_{jk}\} 
 \{\tau_i \cdot \tau_j,\tau_j \cdot \tau_k\}  \\
& + & {{1}\over {4}} 
  [X_{ij},X_{jk}] [\tau_i \cdot \tau_j,\tau_j \cdot \tau_k] \Big),
\end{eqnarray}
\noindent
where

\begin{equation}
X_{ij} = Y(r_{ij}) \sigma_i \cdot \sigma_j + T(r_{ij}) S_{ij} 
\end{equation}
\noindent
is the one--pion exchange operator, $\sigma$ and $\tau$ are the Pauli spin 
and isospin operators, and 
$ S_{ij} =  3 \big[ (\sigma_i \cdot r_{ij})(\sigma_j \cdot r_{ij})
                - \sigma_i \sigma_j \big ] $   
is the  tensor operator. 
$Y(r)$ and $T(r)$ are the Yukawa and tensor functions, respectively, 
associated to the one--pion exchange, as in the two--body potential.    

The repulsive part is taken as  
\begin{equation}
V^{R}_{ijk} = U \sum_{cyc} T^2(r_{ij})  T^2(r_{jk})
\end{equation}
\noindent
The constants A and U in the previous equations can be adjusted to 
reproduce observed nuclear properties. Within this scheme 
Schiavilla {\it et al.} (1986) found $A = - 0.0333$ and $ U = 0.0038$
by fitting properties of light nuclei ($^3$H, $^4$He),
the so called Urbana VII parametrization (UVII) (Carlson et al. 1983).

%%%%%%%%%%%%%%%%%%%%%%%%%%%%%%%%%%%%%%%%%%%%%%%%%%%%%%%%%
% Figure 3
\begin{figure*} 
 \begin{center}
\includegraphics[bb= 60 0 515 719,angle=90,scale=0.5]{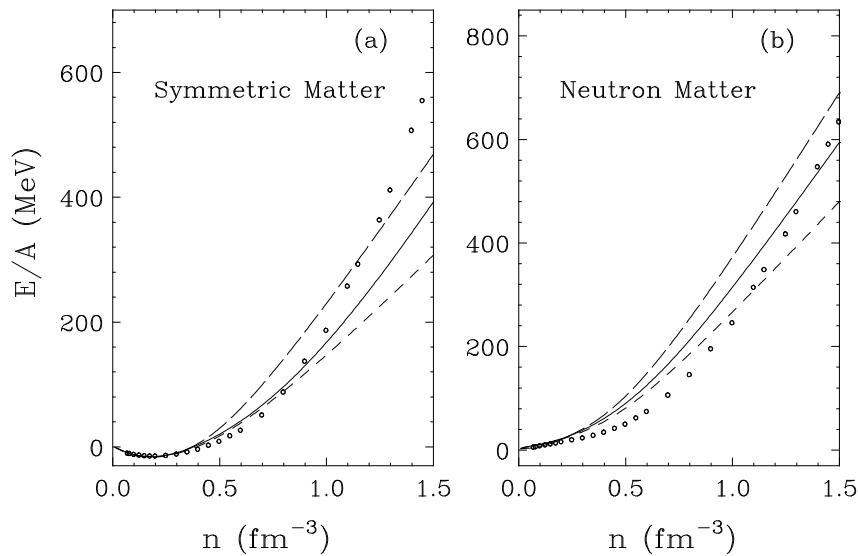}
\end{center}
\caption{The energy per baryon E/A is plotted vs. the 
number density n  for symmetric matter (panel (a)) and 
for neutron matter (panel (b)). 
Several EOS's are shown, {\it i.e.} non-relativistic Brueckner calculations
with three-body forces (solid line, Paris potential; short dashed line, 
AV14 potential) and a relativistic Dirac-Brueckner one (long dashes, DBHF). 
For comparison a variational calculation (with AV14+UVII interaction)  
is also reported (open circles, WFF).}
\label{Fig3}
\end{figure*}
%%%%%%%%%%%%%%%%%%%%%%%%%%%%%%%%%%%%%%%%%%%%%%%%%%%%%%%%%

We introduced the same Urbana three-nucleon model within 
the BHF approach. 
To incorporate the TBF in the Brueckner scheme we followed the method
of Lejeune et al. (1986). The TBF is reduced to an effective two-body
force by averaging on the position of the third particle, assuming
that the probability of having two particles at a given distance
is reduced according to the two-body correlation function. The resulting
effective two-body force is of course density dependent. Further details
will be given elsewhere (Baldo et al. in preparation).
We have adjusted the parameters A and U in order to reproduce closely
the correct saturation point of SNM, since for NS studies this is
an essential requirement, and there is no reason to believe that
TBF be the same as in light nuclei. 
Our values for the TBF parameters are  $A'=-0.0329$ and $U'=0.00361$
both for the Paris and the AV14 potentials.  

The corresponding EOS's obtained using the Paris (so\-lid li\-ne) and
the AV14 potentials (short dashed line) are depicted in Fig. 3
for symmetric (panel (a)) and neutron matter (panel (b)).
The corresponding values are reported in Table 1.
Those EOS's saturate respectively at  

\begin{equation}
n_o = 0.176~fm^{-3},~~~E_o/A = -16.01~MeV ~~~~Paris 
\end{equation}
\noindent
and 

\begin{equation}
n_o = 0.178~fm^{-3},~~~ E_o/A = -16.46~MeV~~~~AV14
\end{equation}

They are characterized by an incompressibility 
$K_{\infty}^{Paris} = 281~MeV $ and $K_{\infty}^{AV14} = 253~MeV $, 
the latter being   
very close to the recent phenomenological estimate of Myers (1996).
In the same figure, we show the EOS obtained within the variational 
many--body approach by WFF when the UVII parametrization (Carlson et al. 1983, 
Schiavilla et al. 1986) is added to the AV14 two--body force
(open circles). 

%%%%%%%%%%%%%%%%%%%%%%%%%%%%%%%%%%%%%%%%%%%%%%%%%%%%%%%%%%%%%%%%%%%%%%%%
% Table A1
\begin{table*}
 \caption{The energy per baryon $E/A$ in MeV is shown vs. the 
number density $n$ in $fm^{-3}$ for symmetric matter and 
neutron matter.
The values are the results of 
a non-relativistic BHF calculation with AV14 and Paris potentials 
with three-body forces.}\label{A1}
\begin{center}
\begin{tabular}{|ccccc|}
\hline
n & ${AV14 + TBF}_{sym}$ & ${Paris + TBF}_{sym}$ & 
${AV14 + TBF}_{neu}$ & ${Paris + TBF}_{neu}$ \\
\hline
0.08   & -11.47   & -10.8  & 8.54   & 9.47    \\
0.16   & -16.32   & -15.88 & 16.2   & 17.  \\
0.2    & -16.25   & -15.74 & 21.    & 22.1  \\
0.3    & -10.78   & -9.4   & 35.6   & 38.8   \\
0.4    & 0.79     &  3.2  & 55.26  & 61.4   \\ 
0.5    & 17.53    &  20.46 & 81.13  & 90.3   \\
0.6    & 38.5     &  41.78  & 112.5    &125.55    \\
0.7    & 62.7     &  67.17  & 147.9     &166.7     \\ 
0.8    & 89.3     &  96.7   & 185.9   &212.66     \\ 
0.9    & 117.67   & 130.22  & 225.7  & 262.3    \\ 
1.     & 147.35   & 167.4   & 266.7    &314.6    \\
1.1    & 178.     & 207.76  & 308.4   & 368.78    \\
1.2    & 209.5    & 250.87  & 350.8   & 424.12   \\
1.3    & 241.58   & 296.27  & 393.5  & 480.3  \\
1.4    & 274.17   & 343.6   & 436.6  & 537.   \\
1.5    & 307.18   & 392.52  & 479.8   & 594.2   \\
\hline
\end{tabular}
\end{center}
\end{table*}
%%%%%%%%%%%%%%%%%%%%%%%%%%%%%%%%%%%%%%%%%%%%%%%%%%%%%%%%%%%%%%%%%%%%%%%%

We have already noticed that both the BHF and the variational
many-body methods produce
a different saturation point already at the two-body level.  
As a consequence, the values of the strengths ($A'$ and $U'$) 
of the attractive 
and repulsive TBF needed to reproduce empirical saturation are different 
with respect to those used by WFF in their work. 
In particular, being $U' < U$, our repulsive TBF is weaker.
Consequently, our EOS's are softer than the WFF at very high 
density (see Fig.3) where the repulsive component of the TBF is dominant. 

In Fig. 3 we plot also the EOS from a recent Dirac-Brueckner calculation 
(DBHF) (Li et al. 1992) with the Bonn--A two--body force 
(long dashed line).      
In the low density region ($n < 0.4~fm^{-3}$), both BHF equations 
of state with TBF and DBHF equations of state are very similar, 
whereas at higher density the DBHF is stiffer.  
The discrepancy between the non-relativistic and relativistic 
calculation of the EOS can be
easily understood by noticing that the DBHF treatment is equivalent
(Baldo et al. 1995) to introduce in the non-relativistic BHF the three-body
force corresponding to the excitation of a nucleon-antinucleon
pair, the so-called Z-diagram (Brown et al. 1987).  The latter is repulsive
at all densities. In BHF treatment, on the contrary, both attractive and 
repulsive three-body forces are introduced, and therefore a softer 
EOS can be expected.  
It should be noticed that, because of the strong repulsion at short
distances, our BHF calculation with three-body force is numerically
stable up to densities $n \simeq 0.8~fm^{-3}$. Moreover, the 
relativistic DBHF calculations are available in literature up to
densities $n~=~0.75~fm^{-3}$ (Li et al., 1992). 
Since the typical densities encountered in the core
of neutron stars are larger, both microscopic calculations need to be  
either extrapolated to slightly higher densities or joined with other
high density equations of state, several of which are  available 
in literature. We decided to perform a numerical extrapolation 
to the needed higher densities. In the extrapolation procedure
we keep checking that the {\it causality condition} is fulfilled, {\it i.e.}

\begin{equation}
c_s/c \equiv \bigg({{dP}\over{d\varepsilon}}\bigg)^{1/2} \leq 1, 
\end{equation}
\noindent
which means that the speed of sound $c_s$ in matter must be lower 
than the speed of light in the vacuum.   
This is a basic trait of any ``realistic'' EOS, regardless the details of the 
interactions among matter constituents or the many--body approach.
Details on the extrapolation procedure are given in Sec.3.

The ratio  $c_s/c$ as a function of 
the number density is reported in Fig. 4, in the case of pure neutron 
matter, for our BHF models (AV14 + TBF(short dashes), Paris + TBF
(solid)), and the DBHF model (long dashes).
WFF model (open circles) violates the  causality condition at  densities 
encountered in the core of neutron stars near the maximum mass 
configuration predicted by that model (Wiringa et al.,1988).

%%%%%%%%%%%%%%%%%%%%%%%%%%%%%%%%%%%%%%%%%%%%%%%%%%%%%%%%%
% Figure 4
\begin{figure} [h]
 \begin{center}
\includegraphics[bb= 60 0 515 669,angle=90,scale=0.6]{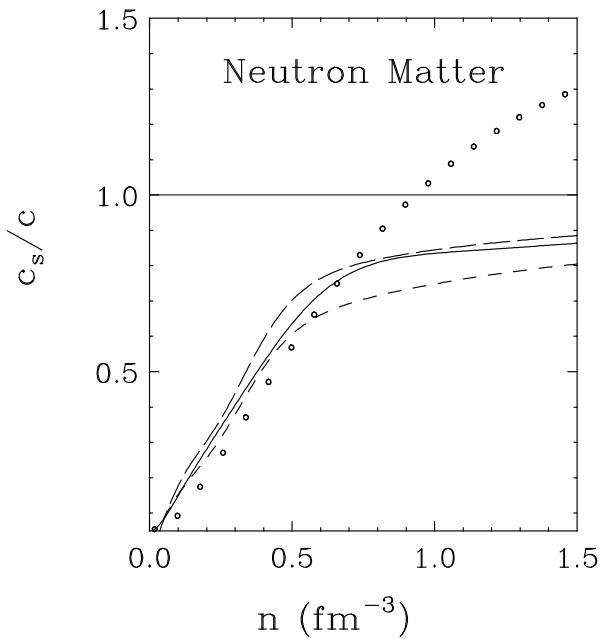}
\end{center}
\caption{The ratio $c_s/c$ is plotted 
as a function of the number density for pure neutron matter.
The notation is the same as in Fig.3.}
    \label{Fig4}
\end{figure}
%%%%%%%%%%%%%%%%%%%%%%%%%%%%%%%%%%%%%%%%%%%%%%%%%%%%%%%%%

%%%%%%%%%%%%%%%%%%%%%%%%%%%%%%%%%%%%%%%%%%%%%%%%%%%
\subsection{Symmetry energy and EOS for $\beta$--stable matter}
%%%%%%%%%%%%%%%%%%%%%%%%%%%%%%%%%%%%%%%%%%%%%%%%%%%

%%%%%%%%%%%%%%%%%%%%%%%%%%%%%%%%%%%%%%%%%%%%%%%%%%%%%%%%%
% Figure 5
\begin{figure*} 
 \begin{center}
\includegraphics[bb= 60 0 515 719,angle=90,scale=0.6]{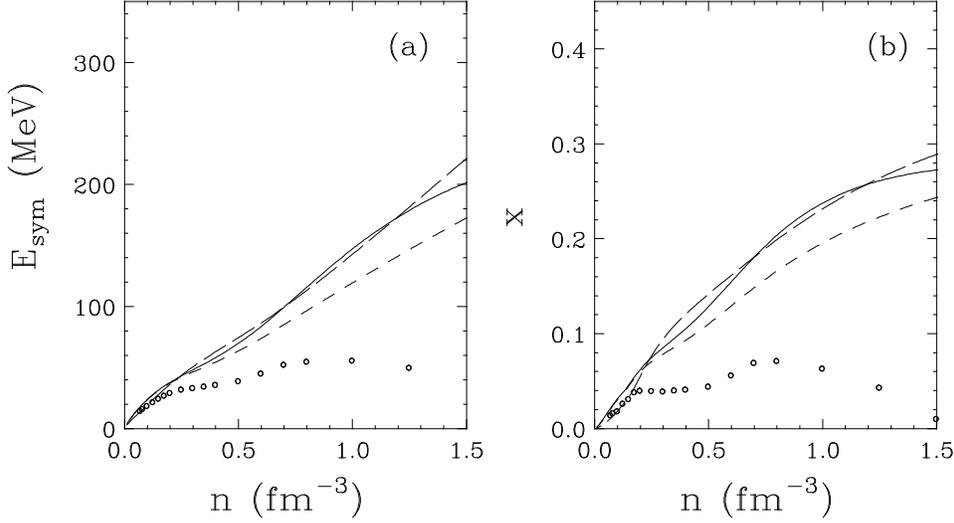}
\end{center}
\caption
{The symmetry energy and the proton fraction
are shown vs. number density respectively in panel (a) and (b).
The notation is the same as in Fig.3.}
\label{Fig5}
\end{figure*}
%%%%%%%%%%%%%%%%%%%%%%%%%%%%%%%%%%%%%%%%%%%%%%%%%%%%%%%%%

The energy per nucleon of asymmetric nuclear matter ($n_n\neq n_p$, being 
$n_n$ and $n_p$ the neutron and proton number densities respectively) 
can be calculated extending the original BBG theory for SNM sketched in 
sect. 2.1. (Bombaci et al. 1994). 
From the energy per nucleon one defines the nuclear symmetry energy as 

\begin{equation}
E_{sym}(n) \equiv {1 \over 2}  {{\partial ^2 E/A }\over{\partial 
                           \beta ^2}}\Big|_{\beta=0}     
\end{equation}
\noindent
where we introduce the {\it asymmetry parameter} 

\begin{equation}
\beta = {{n_n - n_p}\over{n}},
\end{equation}
\noindent
with $ n = n_n + n_p$. 
At the saturation density, $ E_{sym}(n_o)$ corresponds to the symmetry 
coefficient in the Weizs\"acker--Bethe semiempirical mass formula, and its 
value extracted from nuclear systematics lies in the range 28--32~MeV.  

In the present work, we calculate the energy per nucleon of asymmetric 
nuclear matter assuming the so called {\it parabolic approximation}

\begin{equation}
{{E}\over{A}}\big(n,\beta\big) = {{E}\over{A}}\big(n,\beta=0\big) 
                               + E_{sym}(n)\beta^2,   
\end{equation}
\noindent
In this approximation the symmetry energy can be expressed in terms of the 
difference of the energy per particle between neutron ($\beta=1$) and 
symmetric ($\beta=0$) matter:

\begin{equation}
E_{sym}(n) = {{E}\over{A}}\big(n,\beta =1\big) - 
       {{E}\over{A}}\big(n,\beta =0\big) 
\end{equation}
\noindent
The reliability of the parabolic approximation has been checked 
microscopically by Bombaci and Lombardo (1991). 
Then we determine the composition of $\beta$--stable matter, at a given 
nucleon number density, solving the equations for chemical equilibrum (for 
neutrino--free matter, $\mu_{\nu_e} =  \mu_{\nu_\mu} = 0$) 

\begin{equation}
\mu_e = \mu_n - \mu_p \equiv \hat\mu    
\end{equation}

\begin{equation}
\mu_e = \mu_\mu 
\end{equation}
\noindent
and charge neutrality
\begin{equation}
n_p  =  n_e + n_\mu,      
\end{equation}
\noindent
being $\mu_i$ ($i=n,p,e,\mu$) the chemical potentials of star constituents.
Therefore, in the case of neutrino--free matter, the leptonic chemical 
potential  is determined once the difference  
$\hat \mu \equiv \mu_n - \mu_p$ is known.  
 The  latter quantity can be expressed as: 
\begin{equation}
\hat \mu = - {{\partial E/A}\over{\partial x}}\Bigg|_n = 
           2 {{\partial E/A}\over{\partial \beta}}\Bigg|_n  
\end{equation}
\noindent
where $x = n_p/n = (1 - \beta)/2$ is the proton fraction and 
the partial derivatives are taken for constant nucleon number density $n$. 
In eq. (21) we neglect the neutron--proton mass difference.   

In the parabolic approximation (16) for the energy per nucleon of 
asymmetric nuclear matter  one has  

\begin{equation}
\hat \mu = 4 E_{sym}(n) (1-2x)
\end{equation}
\noindent
Therefore the composition of $\beta$-stable matter, and in particular 
the proton fraction $x$, is strongly dependent on the nuclear symmetry energy. 

As it has been recently pointed out by Lattimer et al. (1991), the value of 
the proton fraction in the core of NS  is crucial for the onset of 
direct Urca processes, whose occurrence enhances neutron star cooling rates.

%%%%%%%%%%%%%%%%%%%%%%%%%%%%%%%%%%%%%%%%%%%%%%%%%%%%%%%%%
% Figure 6
\begin{figure} [h]
 \begin{center}
\includegraphics[bb= 60 0 515 669,angle=90,scale=0.6]{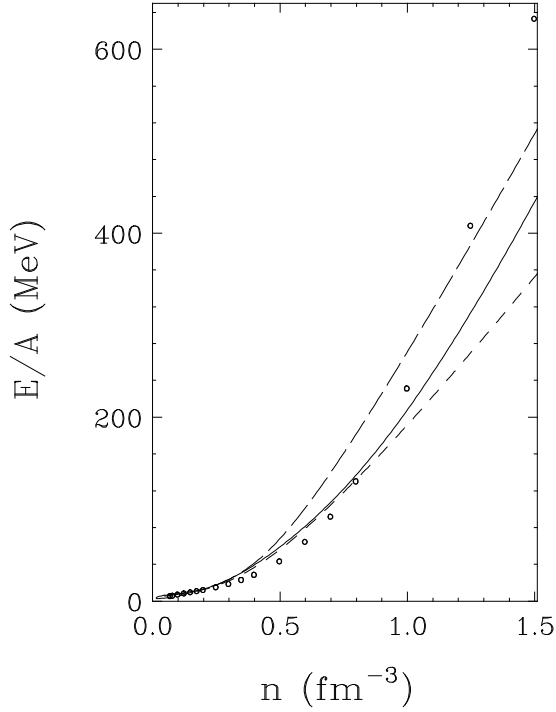}
\end{center}
\caption{The  energy per baryon  E/A  of asymmetric nuclear matter 
in $\beta$-equilibrium (with electrons and muons) is plotted vs. the 
number density $n$.  Different curves are denoted as in Fig.3.}
    \label{Fig6}
\end{figure}
%%%%%%%%%%%%%%%%%%%%%%%%%%%%%%%%%%%%%%%%%%%%%%%%%%%%%%%%%

The values of $E_{sym}$ for the different EOS's are reported in Fig. 5
(panel (a)), together with the corresponding proton fraction (panel (b)). 
The long dashed line represents the relativistic DBHF 
calculation\footnote{In a recent paper (Lee et al., 1997) the authors 
show microscopic calculations of asymmetric nuclear matter in the 
DBHF approach up to densities $n = 1.1~ fm^{-3}$. They obtain a symmetry
energy which saturates with increasing density, at variance with 
our findings. In spite of that the values of the maximum mass 
configuration are similar to those calculated in this work.},
whereas the non--relativistic Brueckner calculations with the
AV14 + TBF and the Paris + TBF models are respectively represented
by a short dashed and a solid line. 
We notice that in both relativistic and non--relativistic 
Brueckner--type calculations, the proton fraction $x$
can exceed the "critical" value 
$x^{Urca} = (11-15)\%$ needed for the occurrence of direct Urca 
processes (Lattimer et al. 1991).  
This is at variance with the WFF variational calculation (open circles), 
which predicts a low absolute value both for the symmetry energy and 
the proton fraction with a slight bend over.   
For the AV14 + TBF model we find $x^{Urca} = 13.9\%$, 
which correspond to a critical density $n^{Urca} = 0.65~fm^{-3}$,
whereas the Paris + TBF model predicts the same critical proton
fraction at density $n^{Urca} = 0.54~fm^{-3}$.
Therefore, BHF neutron stars with a central density higher than
$n^{Urca}$ develop inner cores in which direct Urca 
processes are allowed.       

In Fig.6, we show  the energy per baryon of asymmetric nuclear matter 
in $\beta$-equilibrium,  for the models under consideration.    
We notice that the AV14 and the Paris two-nucleon interactions,
implemented by the same three-body force, produce similar EOS's
up to densities $n \simeq 0.8~fm^{-3}$, whereas the high density 
region is quite different. Therefore we find the same trend as observed
in Fig.2.

  From the energy per baryon of asymmetric nuclear matter in 
$\beta$-equilibrium, we calculate the nuclear contribution $P_{nucl}$ to  
the total pressure of stellar matter using the thermodynamical relation

\begin{equation}
P_{nucl}(n) = n^2 {d~(E/A)\over{dn}} \bigg|_A 
\end{equation}
\noindent
Then the total pressure and total mass density $\rho$ are given by 

\begin{equation}
P = P_{nucl} + P_{lep}   
\end{equation}

\begin{equation}
\rho  =  {1 \over {c^2}} \Big(\varepsilon_{nucl} + \varepsilon_{lep}\Big) 
=  {1 \over {c^2}} \Big(n~E/A  + n~m_N + \varepsilon_{lep}\Big )
\end{equation}
\noindent
being $P_{lep}$ and  $\varepsilon_{lep}$ the leptonic contributions to the 
total pressure and energy density, $m_N$ the nucleon mass and $c$ the speed 
of light in the vacuum.

Our  EOS's for  $\beta$--stable matter (neutrons, protons, electrons and
muons) calculated with 
AV14 (Paris) potential plus Urbana three-body forces are given respectively 
in Table 2 (3). 

%%%%%%%%%%%%%%%%%%%%%%%%%%%%%%%%%%%%%%%%%%%%%%%%%%%%%%%%%
% Table 2
\begin{table}
 \caption{EOS for $\beta$--stable matter obtained in BHF approximation 
using the Argonne AV14 two-body interaction implemented by TBF as 
described in the text. We display the number density $n$ in units of 
$fm^{-3}$, the proton fraction x, the total mass density $\rho$ and the 
total pressure P.}
\begin{tabular}{|cccc|}
\hline
n  & x & $\rho~(10^{14}~g~cm^{-3})$ & $P~(10^{34}~dyn~cm^{-2})$ \\
\hline
0.08  & 0.025 & 1.35 & 0.042  \\
0.16  & 0.048 & 2.7 & 0.244  \\
0.2  & 0.06 & 3.4 & 0.502  \\
0.3  & 0.078 & 5.15 & 1.88  \\
0.4  & 0.093 & 6.98 & 4.62  \\
0.5  & 0.11 & 8.92 & 8.92 \\
0.6  & 0.129 & 10.97 & 14.7  \\
0.7  & 0.148 & 13.13 & 21.87 \\
0.8  & 0.166 & 15.42 & 30.36  \\
0.9  & 0.182 & 17.85 & 40.14 \\
1.  & 0.195 & 20.39 & 51.19  \\
1.1 & 0.207 & 23.07 & 63.5 \\
1.2  & 0.218 & 25.88 & 77.04  \\
1.3 & 0.227 & 28.82 & 91.8 \\
1.4  & 0.236 & 31.88 & 107.78 \\
1.5  & 0.243 & 35.07 & 124.96 \\
\hline
\end{tabular}
\end{table}
%%%%%%%%%%%%%%%%%%%%%%%%%%%%%%%%%%%%%%%%%%%%%%%%%%%%%%

%%%%%%%%%%%%%%%%%%%%%%%%%%%%%%%%%%%%%%%%%%%%%%%%%%%%%%%%%
% Table 3
\begin{table}
\caption{Same as Table 2, but for Paris potential as NN interaction.}
\begin{tabular}{|cccc|}
\hline
n & x & $\rho~(10^{14}~g~cm^{-3})$ & $P~(10^{34}~dyn~cm^{-2})$ \\
\hline
0.08  & 0.025 & 1.35 & 0.039  \\
0.16  & 0.05 & 2.7 & 0.243  \\
0.2  & 0.062 & 3.4 & 0.53  \\
0.3  & 0.085 & 5.17 & 2.07  \\
0.4  & 0.105 & 7. & 4.83  \\
0.5  & 0.128 & 8.95 & 8.95 \\
0.6  & 0.155 & 11.0 & 14.8  \\
0.7  & 0.18 & 13.2 & 23. \\
0.8  & 0.204 & 15.55 & 34.  \\
0.9  & 0.223 & 18.06 & 48. \\
1.  & 0.237 & 20.77 & 64.9  \\
1.1 & 0.25 & 23.67 & 84.53 \\
1.2  & 0.257 & 26.78 & 106.6  \\
1.3 & 0.264 & 30.1 & 131. \\
1.4  & 0.269 & 33.6 & 157.6 \\
1.5  & 0.272 & 37.3 & 186.4 \\
\hline
\end{tabular}
\end{table}
%%%%%%%%%%%%%%%%%%%%%%%%%%%%%%%%%%%%%%%%%%%%%%%%%%%%%%%%%

%%%%%%%%%%%%%%%%%%%%%%%%%%%%%%%%%%%%%%%%%%%%%%%%%%%%%%%%%%%%%%%%%%%%%%%%%
          \section{Neutron star structure }
%%%%%%%%%%%%%%%%%%%%%%%%%%%%%%%%%%%%%%%%%%%%%%%%%%%%%%%%%%%%%%%%%%%%%%%%%
The EOS for $\beta$--stable matter  can be used in the 
Tolman--Oppenheimer--Volkoff (Tolman 1934, Oppenheimer et al. 1939)
equations to compute the neutron star 
mass and radius as a function of the central density. 

We assume that a neutron star is a spherically symmetric distribution of 
mass in hydrostatic equilibrium. We neglect the effects of rotations and 
magnetic fields. Then the equilibrium configurations 
are simply obtained by solving the Tolman-Oppenheimer-Volkoff (TOV) 
equations for the total pressure $P$ and the enclosed mass $m$, 

\begin{equation}
{dP(r)\over{dr}} = -{{G m(r) \rho(r)} \over{r^2}} 
{ {(1 + {P(r)\over {c^2 \rho(r)}})} 
{(1 + {4\pi r^3 P(r)\over {c^2 m(r)}})} \over
{ {(1 - {2G m(r)\over {r c^2 }})} }}
\end{equation}

\begin{equation}
{dm(r)\over{dr}} = 4 \pi r^2 \rho(r)
\end{equation}
\noindent
being $G$ the gravitational constant. 
Starting with a central mass density $\rho(r=0) \equiv \rho_c$,  
we integrate out until the pressure on the surface equals the one 
corresponding to the density of iron.
This gives the stellar radius $R$ and the gravitational mass 
is then 
\begin{equation}
M_G~ \equiv ~ m(R)  = 4\pi \int_0^Rdr~ r^2 \rho(r). 
\end{equation}
\noindent
For the outer part of the neutron star we have used the equations of state
by Feynman-Metropolis-Teller (Feynman et al. 1949) and Baym-Pethick-Sutherland 
(Baym et al. 1971), 
and for the middle-density regime ($0.001~fm^{-3}<n<0.08~fm^{-3}$) we use the 
results of Negele and Vautherin (Negele et al. 1973). 
In the high-density part ($n > 0.08~fm^{-3}$) we use alternatively 
the four EOS's  discussed above.   
The results are reported in Fig. 7. 
We display the gravitational mass $M_G$, in units of the solar mass 
$M_{\sun}$ ($M_{\sun} = 1.99~10^{33}$ g),  as a function 
of the radius R (panel (a)) and the central number density $n_c$
(panel (b)). The notation is the same as in the previous figures.

As expected, the stiffest EOS within the Brueckner scheme (DBHF) 
we used in the present 
calculation gives higher maximum mass and lower central density with 
respect to the non-relativistic Brueckner models. 

The properties of the maximum mass configuration, for each EOS, are 
summarized in table 4, whereas in table 5 the properties of neutron
stars with gravitational mass $M_G = ~1.4~ M_{\sun}$ are reported.
However, the properties of each configuration extracted by solving
the TOV equations are slightly dependent on the extrapolation parameters
we used. This holds for the DBHF equation of state and the
Paris + TBF model, while in the case of Argonne AV14 + TBF model 
no seizable dependence was found. For the BHF calculation with only 
two-body forces no
extrapolation is needed. In any case the dependence on the
extrapolation procedure turns out to be rather weak. We have used
a fractional polynomial extrapolation of the form 

\begin{equation}
{\frac{E} {A}} = {\frac {Q^\sigma (n)} {1 + b~n^\sigma}}
\end{equation}
\noindent
where the coefficients of the polynomial  $Q$ are fitted to reproduce the 
EOS up to the maximum calculated density $n$, and the coefficient $b$ ensures 
that for very large values of n the energy per baryon does not diverge.   
Different values for the polynomial degree
$\sigma$ can be used and different extrapolated EOS are in general obtained. 
However we found that the critical mass, radius and central densities 
do not vary more than $5 \%$  with $\sigma = 3,~4$ and
$b$ ranging from $10^{-2}$ to $10^{-3}$ for the case of the Paris potential +
TBF model calculations, while in all other cases the variations can be 
considered negligeable.

%%%%%%%%%%%%%%%%%%%%%%%%%%%%%%%%%%%%%%%%%%%%%%%%%%%%%%%%%
% Figure 7
\begin{figure*} 
 \begin{center}
\includegraphics[bb= 50 0 515 719,angle=90,scale=0.5]{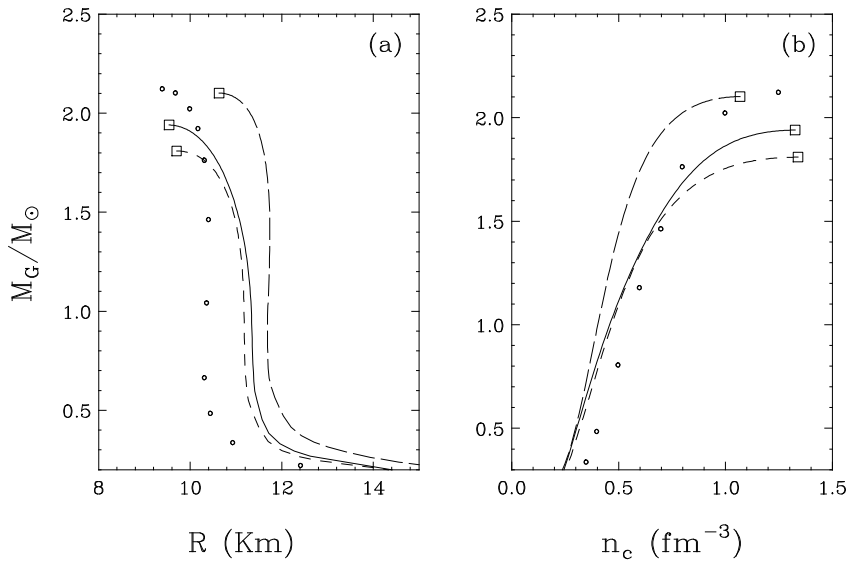}
\end{center}
\caption
{The gravitational mass $M_G$, expressed in units
of the solar mass $M_{\sun}$, is displayed vs. radius R (panel (a))
and the central number density $n_c$ (panel (b)). 
The notation is the same as in Fig.3. The squares represent the values 
of the limiting configuration.}
\label{Fig7}
\end{figure*}
%%%%%%%%%%%%%%%%%%%%%%%%%%%%%%%%%%%%%%%%%%%%%%%%%%%%%%%%%

As one can see from Table 4, the values
of the limiting mass configuration are strongly modified by 
the presence of the nuclear TBF in the equation of state,
in particular by their repulsive contribution which dominates at high 
densities. However, the non-relativistic BHF calculations implemented
by three-body force do not strongly depend on the two-body
interaction, and predict very close limiting values of the maximum mass
(within the numerical uncertainties due to the extrapolation).     

The difference between BHF and WFF neutron stars reflects the discrepancy 
already noticed for the EOS and mainly for the symmetry energy. 
This point will be discussed in more details in a forthcoming paper
(Baldo et al., in preparation).

Our calculated BHF configurations with three-body forces are consistent 
with  the measured masses of X--ray pulsars and radio pulsars 
(van Kerkwijk et al. 1995).

%%%%%%%%%%%%%%%%%%%%%%%%%%%%%%%%%%%%%%%%%%%%%%%%%%%%%%%%%%%%%%%%%%%%%%%%
% Table 4
\begin{table}
 \caption{Properties of  the maximum mass configuration obtained for different 
          equations of state: 
$M_G$ is the gravitational (maximum) mass,
R is the corresponding radius, $n_c$ the central number 
density and $x_c$ the central proton fraction.}
\begin{tabular}{|ccccc|}
\hline
EOS & $M_G/M_{\sun}$ & R(km) & $n_c(fm^{-3})$ & $x_c$ \\
\hline
      AV14  & 1.5 & 8.1 & 1.9 & 0.17 \\
      Paris & 1.66 & 8. & 1.88 & 0.26 \\
      AV14 + TBF & 1.8 & 9.7 & 1.34 & 0.23 \\
      Paris + TBF & 1.94 & 9.54 & 1.33 & 0.265 \\
      DBHF  & 2.1     & 10.6  &  1.07 & 0.24  \\
      WFF   & 2.130     &  9.40   &  1.25  & 0.045  \\
\hline
\end{tabular}
\end{table}
%%%%%%%%%%%%%%%%%%%%%%%%%%%%%%%%%%%%%%%%%%%%%%%%%%%%%%%%%%%%%%%%%%%%%%%%

%%%%%%%%%%%%%%%%%%%%%%%%%%%%%%%%%%%%%%%%%%%%%%%%%%%%%%%%%%%%%%%%%%%%%%%%
% Table 5
\begin{table}
 \caption{Properties of  neutron stars with $M_G = 1.4~M_{\sun}$ } 
\begin{tabular}{|cccc|}
\hline
 EOS & R(km) & $n_c(fm^{-3})$  & $x_c$ \\
\hline
      AV14 & 8.92  &  1.27   & 0.12   \\
      Paris & 9.3 &  1.08  & 0.178  \\
      AV14 + TBF  & 11. & 0.65 & 0.139 \\
      Paris + TBF & 11.09  & 0.64 & 0.165 \\
      DBHF  & 11.74 & 0.49 & 0.14   \\
      WFF   & 10.41& 0.66  &  0.0658  \\
\hline
\end{tabular}
\end{table}
%%%%%%%%%%%%%%%%%%%%%%%%%%%%%%%%%%%%%%%%%%%%%%%%%%%%%%%%%%%%%%%%%%%%%%%%

%%%%%%%%%%%%%%%%%%%%%%%%%%%%%%%%%%%%%%%%%%%%%%%%%%%%%%%%%%%%%%%%%%%%%%%%%
           \section{Conclusions}
%%%%%%%%%%%%%%%%%%%%%%%%%%%%%%%%%%%%%%%%%%%%%%%%%%%%%%%%%%%%%%%%%%%%%%%%%

In conclusion, we computed some properties of NS's on the basis of a 
microscopic EOS obtained in the framework of BBG many--body theory with 
two-- and three--body nuclear interactions. Our Brueckner-Hartree-Fock 
EOS with three-body forces satisfies the general 
physical requirements (points i--iv) discussed in the introduction. 
This is the main feature which distinguishes our EOS's with respect to 
other microscopic non--relativistic EOS (Wiringa et al. 1988, Engvik et al.
1994, 1996).    
Two different two-body potentials have been used, the Argonne AV14 and the
Paris interactions. The latter turns out to produce a slightly stiffer
EOS already at two-body level. The same three-body force, within the 
Urbana model, has been added to both two-body interactions, and this
produces similar saturation points close to the empirical one. The BHF
calculations with three-body forces show again that the Paris potential gives
an EOS stiffer than the one from the Argonne AV14 and a larger symmetry
energy as a function of density. As a consequence the proton fraction is
larger for the Paris and the Urca process onset appears at lower density. 

The calculated maximum mass is in agreement with observed NS 
masses (van Kerkwijk et al. 1995). However the limiting mass, radius and 
central density are relatively close to each other for the two interactions.
In particular the critical masses differ by no more than $8 \%$, while the
radius and central density are essentially equal. On the contrary all 
these three critical physical parameters are appreciably different from the 
ones derived from the Dirac-Brueckner approach (Li et al. 1992) as well
as from the variational results. The latter displays also a rather different 
trend of the symmetry energy. 

For the Argonne AV14 (Paris) potential the BHF approach with 
three-body forces predicts a lower limit for the mass above which 
neutron stars can support the direct Urca process, 
$M^{Urca} = 1.4 M_{\sun}$ ($1.24 M_{\sun}$). 
The stars with mass larger than 
this value cool very rapidly or not depending on the properties of nuclear  
superluidity (values of the superfluid gaps, critical temperatures, density 
ranges for the superfluid transition)(Page et al. 1992, Baldo et al. 1995).
The EOS's developed in this paper offer the possibility for a selfconsistent 
microscopic calculation for 
both the neutron star structure, and nuclear superfluid properties  
within the same many--body approach and with the same nuclear interaction. 

%\appendix 
%\section{}

%Here we report the values of the energy per baryon $E/A$ 
%vs. baryon number $n$  found in the Brueckner-Hartree-Fock 
%approach, using the AV14 and the Paris potentials
%implemented by Urbana model for the three-body force.
%We show the values both for symmetric and neutron matter.

\end{document}